\documentclass[preprint,superscriptaddress,aip,onecolumn]{revtex4-1}
\usepackage{graphicx}
\usepackage{dcolumn}
\usepackage{bm}
\usepackage{amsmath}
\usepackage[english]{babel}
\usepackage{hyperref}
\usepackage{color}

\begin{document}

\title{Spontaneous emission inhibition of telecom-band quantum disks inside single nanowire on different substrates}
\author{Muhammad Danang Birowosuto}
\affiliation{NTT Basic Research Laboratories, NTT Corporation, 3-1 Morinosato Wakamiya, Atsugi, Kanagawa 243-0198, Japan}
\affiliation{NTT Nanophotonics Center, NTT Corporation, 3-1 Morinosato Wakamiya, Atsugi, Kanagawa 243-0198, Japan}
\author{Guoqiang Zhang}
\affiliation{NTT Basic Research Laboratories, NTT Corporation, 3-1 Morinosato Wakamiya, Atsugi, Kanagawa 243-0198, Japan}
\author{Atsushi Yokoo}
\affiliation{NTT Basic Research Laboratories, NTT Corporation, 3-1 Morinosato Wakamiya, Atsugi, Kanagawa 243-0198, Japan}
\affiliation{NTT Nanophotonics Center, NTT Corporation, 3-1 Morinosato Wakamiya, Atsugi, Kanagawa 243-0198, Japan}
\author{Masato Takiguchi}
\affiliation{NTT Basic Research Laboratories, NTT Corporation, 3-1 Morinosato Wakamiya, Atsugi, Kanagawa 243-0198, Japan}
\affiliation{NTT Nanophotonics Center, NTT Corporation, 3-1 Morinosato Wakamiya, Atsugi, Kanagawa 243-0198, Japan}
\author{Masaya Notomi}
\affiliation{NTT Basic Research Laboratories, NTT Corporation, 3-1 Morinosato Wakamiya, Atsugi, Kanagawa 243-0198, Japan}
\affiliation{NTT Nanophotonics Center, NTT Corporation, 3-1 Morinosato Wakamiya, Atsugi, Kanagawa 243-0198, Japan}
\date{\today}

\baselineskip24pt
\begin{abstract}
We investigate the inhibited spontaneous emission of telecom-band InAs quantum disks (Qdisks) in InP nanowires (NWs). We have evaluated how the inhibition is affected by different disk diameter and thickness. We also compared the inhibition in standing InP NWs and those NWs laying on silica (SiO$_{2}$), and silicon (Si) substrates. We found that the inhibition is altered when we put the NW on the high-refractive-index materials of Si. Experimentally, the inhibition factor $\zeta$ of the Qdisk emission at 1,500 nm decreases from 4.6 to 2.5 for NW on SiO$_{2}$ and Si substrates, respectively. Those inhibitions are even much smaller than that of 6.4 of the standing NW. The inhibition factors well agree with those calculated from the coupling of the Qdisk to the fundamental guided mode and the continuum of radiative modes. Our observation can be useful for the integration of the NW as light sources in the photonic nanodevices.
\end{abstract}

\maketitle

\section{Introduction}
The spontaneous emission rate of quantum emitters is known to depend on its nanoscale environment, which is defined by the local density of states (LDOS) \cite{Sprik1996}. This effect has been studied theoretically and experimentally with a wide range of systems, such as reflecting interfaces \cite{Drexhage1970,CPS1978}, photonic crystals \cite{Yablonovitch1987,SajeevJohn1987,Lodahl2004,Noda2005em} and nanocavities \cite{Englund2005,Birowosuto2012}, disordered photonic materials \cite{Birowosuto2010}, and plasmonic nanoantennae \cite{Farahani2005}.

{Semiconductor nanowires (NWs) have promising applications in photonics as the tiny structure embedded with an active material will be a perfect candidate for the smallest devices. As examples, single NWs laying on a plasmonic metal film \cite{Oulton2009Nature,LuYujung2012} and on a dielectric substrate \cite{Jagadish2013,Mayer2013} demonstrated a lasing. The single NW also can be used as a movable cavity in a two-dimensional Si photonic crystal providing a significant degree of flexibility in integrated photonics \cite{Birowosuto2014}.}

Emitter embedded NWs, e.g. quantum dot-NWs, are promising photonic systems since their spontaneous emission can be controlled efficiently \cite{Claudon2010,Bleuse2011,Bulgarini2012}. In this NW, depending on the ratio of the NW diameter ($d$) and the emission wavelength ($\lambda$), emitted photons in the standing NW may funnel efficiently into the fundamental waveguide mode confined in NW. This results in a very directional emission. By contrast, at small $d/\lambda$ ratio, the emission is coupled with a broad continuum of non-guided radiative modes with a result of a strong emission rate inhibition \cite{Bleuse2011,Bulgarini2012}. This inhibition is comparable with the photonic band gap structures \cite{Lodahl2004,Noda2005em}. However, there is no investigation on the spontaneous emission inhibition of the NW horizontally laying on the simplest nanophotonic system, substrates. For a point emitter close to a substrate between two media with different refractive indices, the LDOS is modified by the interference of the emitted and reflected light \cite{Drexhage1970,CPS1978}. In the case of the NW, this interference may alter the spontaneous emission inhibition in the NW.

In this paper, we investigate the spontaneous emission in the telecom band (1,200 - 1,500 nm) of InAs quantum disks (Qdisk) embedded in InP NW as standing and laying on different substrates e.g. SiO$_{2}$ and Si. We use time-resolved photoluminescence (PL) experiments for the spontaneous emission of four or five Qdisks in single NW. Unlike the previous experiments in which they measured randomly distributed single quantum dots in each single NW with various $d$ \cite{Bleuse2011,Bulgarini2012}, here we directly compare the emission rate of multiple Qdisks in one single high-quality NW with no strong variations from one sample to the other sample while controlling the inhibition by growing different thickness $t$ and $d$ of Qdisks. We monitor the Qdisk position in single NW through the PL scan and assign the emission wavelength of specific $t$ and $d$ of Qdisks. {We demonstrate that the emission rates of the Qdisks in the single NW laying on the SiO$_{2}$ still have inhibitions, which are almost the same as those of the standing NW}. However, the inhibition effect decreases as we put the NW on Si substrate. We explain that the decrease of the inhibition is due to the leaking of the guided mode and the continuum of radiative modes as the total internal reflection is broken. Our findings may have implications for the integration of the NW with dielectric substrates, which may find strong interests for the applications in photonic nanodevices \cite{Birowosuto2012b}.

\section{Theory and simulation}
In this section, we will calculate the emission rate of the embedded dipole inside the NW and its relation to the mode confinement as previously already demonstrated in Ref. \cite{Bleuse2011}. However, in our investigation, we extend the concept of Ref. \cite{Bleuse2011} for the NW laying on a dielectric substrate. Here we perform finite-element simulation with two-dimensional model to obtain the parameters for the calculation.

In principle, the normalized total emission rate $P$ of a radial dipole embedded inside a single NW can be separated into two contributions:
\begin{eqnarray}
P = \Gamma_{NW}/\Gamma_{bulk} = P_{m}+P_{c}
\label{decayeq}
\end{eqnarray}
where $\Gamma_{NW}$ and $\Gamma_{bulk}$ are the emission rates of the dipole inside the NW and the bulk InP, respectively. $P_{m}$ and $P_{c}$ are the normalized spontaneous emission rate into the fundamental waveguide mode HE$_{11}$ and into other radiation modes, including a continuum of radiative modes, respectively. {$P$ can be assigned as a measure for the enhancement ($P > 1$) \cite{Purcell1946} or the inhibition ($P < 1$) \cite{Kleppner1981}. For the enhancement, $P$ is similar with the concept of the Purcell factor for a microcavity, which is derived from the LDOS. It describes the spontaneous emission rate in a weak coupling regime \cite{Sauvan2013,vanVlack2012}. However, in our case, the Purcell factor depends only on the waveguide properties and constitutes a figure of merit of the spontaneous emission rate enhancement in a waveguide. In this paper, we will focus on the inhibition of the spontaneous emission rate and later we will assign an inhibition factor $\zeta = 1/P$ to specify the strength of the inhibition}. First, we investigate $P_{m}$ as $P_{m}$ may strongly contribute to $P$ \cite{Friedler2009}. We calculate $P_{m}$ in an infinitely long NW inside air with a cylindrical section of $d$ of 242 nm and a refractive index $n_{NW}$ of InP of 3.17 using \cite{Bleuse2011}:
\begin{eqnarray}
P_{m} = 2 \times \frac{3}{8{\pi}}\frac{(\lambda/n)^2}{S_{eff}}\frac{n_{g}}{n_{NW}}
\label{Purcelleq}
\end{eqnarray}
where $\lambda$ and $n_{g}$ are the vacuum wavelength and the group index, respectively. {In this analytical model, $S_{eff}$ is attributed to the effective NW cross-sectional area.} It describes the mode lateral confinement of the fundamental guided mode HE$_{11}$, which is defined as following:
\begin{eqnarray}
S_{eff} = {\int}^{r_{c}}_{0}{\int}^{2\pi}_{0}\frac{n_{NW}(\textbf{r})^{2}|\textbf{E}(\textbf{r},\theta)|^{2}d\textbf{r}d\theta}{\max[n_{NW}(\textbf{r})^{2}|\textbf{E}(\textbf{r},\theta)|^{2}]}
\label{surfaceeq}
\end{eqnarray}
where $|\textbf{E}(\textbf{r},\theta)|^{2}$ is the electric field intensity inside of the NW { while $r$ and $\theta$ are required for this polar coordinate system}. For an infinite-length NW inside air, the electric field distribution is relatively simple \cite{Bleuse2011}, {see Appendix A1}. For the NW laying on a substrate, the reflection of the fundamental mode on the substrate may contribute to the distribution of the electric field. In fact, the same problem also happens for a finite-length NW standing on a substrate. Friedler \textit{et al}. used a {semi-analytical} Fabry-Perot model with a set of reflection coefficients for understanding the impact of the reflections on the finite-length NW ends to $P_{m}$ \cite{Friedler2009}. In our case, we use the well-known model of a dipole near an interface \cite{Drexhage1970,CPS1978} to understand the impact of the reflection to the emission of the NW laying on the substrate. Here we should consider two component dipole orientations, perpendicular ($\perp$) and parallel ($\parallel$) to the substrate.  Then, we formulate the electric field inside the NW laying on the substrate $\textbf{E}_{int}^{\perp,\parallel}(\textbf{r},\theta)$ as following:
\begin{equation}
\textbf{E}_{int}^{\perp}(\textbf{r},\theta) = \textbf{E}_{NW}(\textbf{r},\theta){\left[1-\left(\int_{0}^{\infty}R^{\parallel}e^{-2l_{NW}k_{NW}h}\frac{u^{3}}{l_{NW}}du\right)\right]^{\frac{1}{2}}}
\label{fieldsubstrate1}
\end{equation}
\begin{equation}
\textbf{E}_{int}^{\parallel}(\textbf{r},\theta) = \textbf{E}_{NW}(\textbf{r},\theta){\left[1+\left(\int_{0}^{\infty}[(1-u^{2})R^{\parallel}+R^{\perp}]e^{-2l_{NW}k_{NW}h}\frac{u}{2l_{NW}}du\right)\right]^{\frac{1}{2}}}
\label{fieldsubstrate2}
\end{equation}
$\textbf{E}_{NW}$ is the electric field inside the NW without the presence of a substrate. $R^{\parallel}$ and $R^{\perp}$ are the Fresnel reflection coefficients for p-polarized and s-polarized, respectively, and they can be written as following:
\begin{equation}
R^{\parallel} = \frac{l_{int}-n_{int}^{2}l_{air}}{l_{int}+n_{int}^{2}l_{air}}\\
\label{Fresnelcoef1}
\end{equation}
\begin{equation}
R^{\perp} = \frac{l_{air}-l_{int}}{l_{air}+l_{int}}
\label{Fresnelcoef2}
\end{equation}
where $l_{air} \equiv -i((1-u^{2})^{1/2}$, $l_{int} \equiv -i((n_{int})^{2}-u^{2})^{1/2}$, and $l_{NW} \equiv -i((n_{NW})^{2}-u^{2})^{1/2}$. {$n_{int}$, $k_{NW}$} and $h$ are {the refractive index of the interface, the propagation constant of the NW}, and the distance of the dipole to the substrate, respectively. On solving $P_{m}$ in Eq. \ref{Purcelleq}, we substitute $|\textbf{E}(\textbf{r},\theta)|^{2}$ in Eq. \ref{surfaceeq} with $|\textbf{E}_{int}(\textbf{r},\theta)|^{2}$ where $|\textbf{E}_{int}(\textbf{r},\theta)|^{2} = 1/3|\textbf{E}_{int}^{\perp}(\textbf{r},\theta)|^{2} + 2/3|\textbf{E}_{int}^{\parallel}(\textbf{r},\theta)|^{2}$ is assumed for the NW with an isotropic emission polarization \cite{CPS1978}. {For the substrate refractive index which is almost the same as that for the air, both reflection coefficients in Eqs. \ref{Fresnelcoef1} and \ref{Fresnelcoef2} will be zero ($R^{\parallel} = R^{\perp} \approx 0$) and therefore  $|\textbf{E}_{int}(\textbf{r},\theta)|^{2} \approx |\textbf{E}_{NW}(\textbf{r},\theta)|^{2}$ and $P_{m}$ for the NW laying on the substrate can be simplified as that for the NW inside the air $P_{m}^{subs} \approx P_{m}^{NW}$}. This works for all dipole position inside the NW. Additionally, if the substrate is a perfect mirror ($R^{\parallel} = R^{\perp} =-1$) and the dipole inside the NW at the boundary with the substrate $h \rightarrow 0$, $|\textbf{E}_{int}^{\perp}(\textbf{r},\theta)|^{2} \approx 2|E_{NW}(\textbf{r},\theta)|^{2}$ is strongly enhanced while $|\textbf{E}_{int}^{\parallel}(\textbf{r},\theta)|^{2} \approx 1/2|E_{NW}(\textbf{r},\theta)|^{2}$ is suppressed. This condition for a perfect mirror and $h \rightarrow 0$ again makes $|\textbf{E}_{int}(\textbf{r},\theta)|^{2} \approx |\textbf{E}_{NW}(\textbf{r},\theta)|^{2}$ and $P_{m}^{subs} \approx P_{m}^{NW}$. After understanding this substrate impact to $P_{m}$, we subsequently focus with the substrate impact to $P_{c}$.

{The emission rate $P_{c}$ caused by the coupling of the emitter to the continuum of radiative modes for the standing NW inside the air was previously discussed in Ref. \cite{Bleuse2011}. This coupling is affected by the screening of the radiative modes as the NW has a high-refractive index \cite{Bleuse2011,Friedler2009}. However, the presence of the substrate in the vicinity of the NW will have an impact to the screening of radiative modes. But in comparison with $P_{m}$, $P_{c}$ in small $d$ NW with the presence of the substrate ($P_{c} \sim n_{NW}^{-5}{[1-(\int_{0}^{\infty}R^{\parallel}e^{-2l_{NW}k_{NW}h}\frac{u^{3}}{l_{NW}}du)]}$) still indicates an emission rate smaller than $P_{m}$ \cite{Bleuse2011,Friedler2009}.}

\begin{figure}[htbp]
\centering\includegraphics[width=12cm]{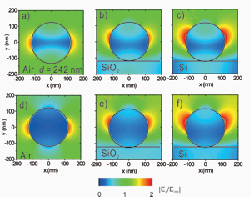}
\caption{Screening of incident plane wave by InP NWs (diameter $d = 242 \pm 7$ nm) inside the air and on different substrates with wavelengths $\lambda =$ 1,200 nm ((a), (b), and (c)) and $\lambda =$ 1500 nm ((d), (e), and (f)). Figures are color maps of $|E_{x}|/E_{inc}$ where $E_{x}$ and $E_{inc}$ are the amplitude of the electrical field along $x$ and that of the incident field, respectively.}
\label{Fig1}
\end{figure}

To investigate the effect of the substrate to the screening of the radiative modes in details, we study the simple problem of the screening of a single plane wave whose wave vector and polarization are perpendicular to the NW axis. Fig. \ref{Fig1} exhibits the field maps for the infinite-length NW with $d$ of 242 nm and two $\lambda$ of 1,200 nm (a,b, and c) and 1,500 nm (d ,e, and f). We simulate three different NW configurations which consist of standing NW (inside the air), NW laying on a SiO$_{2}$, and Si substrates, respectively. Here we use the substrate refractive index for SiO$_{2}$ and Si of 1.53 and 3.49, respectively.

First, we compare two different $d/\lambda$ of the standing NW (\ref{Fig1}a and d for $d/\lambda$ = 0.20 and 0.16, respectively). The field amplitude for $d/\lambda$ = 0.16 reaches minimum and it is uniform inside NW. {When $d/\lambda$ = 0.2, the field becomes strongly dependent on the position inside the NW and the screening effect is not maximized at the NW axis.} This effect is consistent with the previous study \cite{Bleuse2011,Friedler2009,Katsenelenbaum1949}and the field maps also reproduce those observed in Ref. \cite{Bleuse2011}. Then, we investigate the effect of the substrate to the NW. For $d/\lambda$ = 0.2, there is no variation of the average of the field amplitude distributions for the standing NW, the NW on SiO$_{2}$, and the NW on Si substrates (Fig. \ref{Fig1}a, b, and c). The maximum screening effect is only slightly pulled to the direction of the substrate when the refractive index of the substrate increases. However, this is not the case for $d/\lambda$ = 0.16 (Fig. \ref{Fig1}d, e, and f). First, the field amplitude of the NW is uniformly suppressed for the standing NW. When we increase the refractive index of the substrate, the field amplitude in the some cross-sectional parts of the NW increases altering the spontaneous emission inhibition inside the NW.

Using $P_{m}$, {$P_{c}$}, and Eq. \ref{decayeq}, we calculated $P$. First, we obtained $P$ of standing InP NW inside air and $P$ of InP NW laying on SiO$_{2}$ substrate as a function of $d/\lambda$. The results are respectively shown as orange dashed and orange solid lines in Fig \ref{Fig5}c and it is quite similar with previous calculation for GaAs (black dashed lines) \cite{Bleuse2011}. However, the inhibition for both InP NW at $d/\lambda <$ 0.20 is smaller than that of GaAs NW as refractive index of InP in our calculation (3.17) is smaller than that of GaAs (3.45). A plot of $P$ as a function of substrate refractive index for two $\lambda$ of 1,200 and 1,500 nm is also shown in Fig. \ref{Fig4}d. We observe that the increase of emission rate with the refractive index for $\lambda$ = 1,500 nm is larger than that for $\lambda$ = 1,200 nm. On one hand, the inhibition factor {$\zeta$} at 1,500 nm decreases from 6.6 to 2.1 for the standing NW and the NW on Si substrate, respectively. On the other hand, {$\zeta$} at 1,200 nm only decreases from 1.5 to 1.2, respectively.

\section{Experiments}
In our experiments, we used a {standard} microPL setup. {We performed a free space technique with the excitation path and the emission collection from the top using a 0.42 and 0.50-numerical-aperture near-infrared microscope objective \cite{Birowosuto2012}}. {The size of the excitation beam spot} was set between $\sim$2 and $\sim$4 $\mu$m while the emission is first filtered by a dichroic beam mirror. The imaging parts contained a visible-near infrared CCD camera and a short wave infrared InGaAs camera with a {long-wavelength-pass filter} (cut off wavelength of 1,200 nm) for the microscope image and the PL image, respectively. To measure the PL spectrum, we coupled the emission into the multimode fiber and directed it to a grating spectrometer with a cooled InGaAs array. For the time-resolved emission, we filtered the emission with a band pass filter ranging from 1,200 to 1,550 nm and directed it to the NbN superconducting single photon detector (SSPD).

For the samples, we synthesized the InAs/InP NWs in a metalorganic vapor phase epitaxy (MOVPE) system using the {self-catalyst vapour-liquid-solid} (VLS) mode \cite{Zhang2013a,Zhang2013b} without the use of the gold particles. The NW growth starts from the {indium} particle resulting high purity and high quality NWs \cite{Zhang2013a,Zhang2013b}. Those NWs are vertically grown on the substrate based on the epitaxial condition between the NW and the substrate. For the emitters, we designed the multi-stacked InAs/InP heterostructure NWs with very thin InAs Qdisk \cite{Tateno2012,Zhang2013b}. For the experiments, we grew InAs/InP NWs with two different structures, e.g. finite cylindrical and tapered NWs, which later we will assign as samples A and B, respectively.

\begin{figure}[htbp]
\centering\includegraphics[width=9cm]{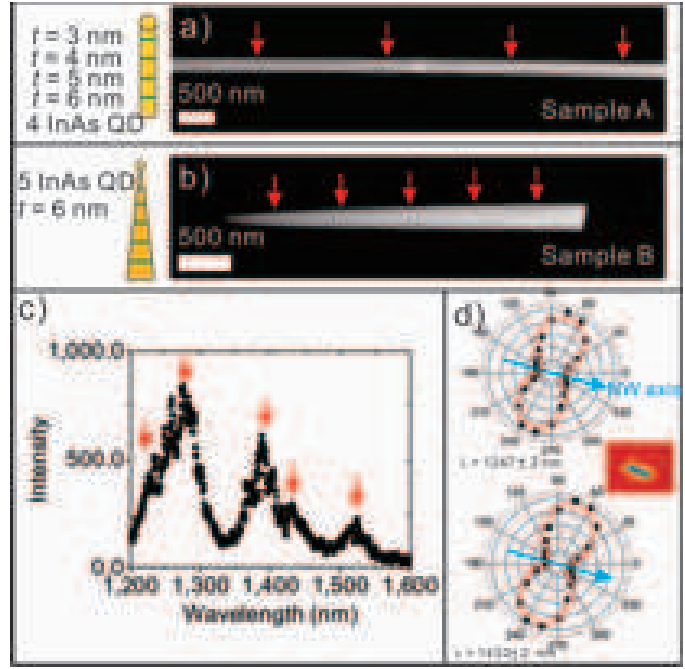}
\caption{(a,b) TEM images of two fabricated InP NWs embedded with multiple InAs Qdisks. The cartoons illustrates the finite cylindrical (sample A) and tapered (sample B) NWs embedded with Qdisks and their thickness $t$ variations. Four different-$t$ and five different-$d$ Qdisks are indicated by red arrows in (a) and (b), respectively. (c) Typical photoluminescence (PL) spectra from single NW, which emission peaks shown by red arrows. 
(d) Azimuthal dependence of the integrated intensity of linearly polarized PL bands for $\lambda$ = 1247 $\pm$ 2 nm and $\lambda$ = 1433 $\pm$ 2 nm. The inset shows the false-color microscope image of the NW (sample B). PL spectra were measured using a CW diode laser excitation at 640 nm wavelength and 4K.}
\label{Fig2}
\end{figure}

The transmission electron microscope (TEM) images and the designs of sample A and B are shown in Fig. \ref{Fig2}a and \ref{Fig2}b, respectively. For sample A, we synthesized InP NWs with $d$ = 242 $\pm$ 11 nm and four InAs Qdisks, in which $t$ was tuned by the variation growth time of 0.5, 1, 1.5, and 2 ns. These growth times yield 2.7, 3.7, 5.5, and 5.7 nm-thick InAs Qdisks. For sample B, $t$ of five InAs Qdisks was kept for 6.0 nm but $d$ was changed from 189 $\pm$ 8 nm to 305 $\pm$ 12 nm through the tapering growth. The typical PL spectrum of InAs (sample B) is shown in Fig. \ref{Fig2}c. As shown by red arrows, we attributed five major emission peaks to the emission from InAs Qdisks. The polarization study of the emission from the Qdisks (sample B) is shown in Fig. \ref{Fig2}d. Most of Qdisks emissions are polarized perpendicular to the axis of the NW. This polarization guarantees an optimal coupling to the guided mode and an efficient screening of the other modes.

\begin{figure}[htbp]
\centering\includegraphics[width=10cm]{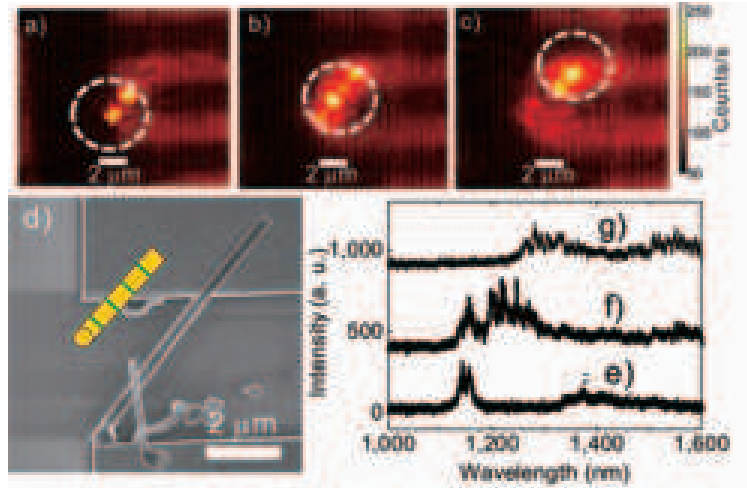}
\caption{(a,b,c) PL images of Qdisks in single finite cylindrical NW (sample A) on gold substrate with different positions of the excitation spot. The excitation beam is shown by the white-dashed circles. (d) SEM of image of the corresponding NW and the cartoon illustrates the direction of the NW. (e,f,g) PL spectra of the Qdisks for different excitation-spot positions. Spectra in (e),(f), and (g) correspond to the images in (a), (b), and (c), respectively. Measurements were performed using a CW diode laser excitation at 640 nm wavelength and 4K.}
\label{Fig3}
\end{figure}

\begin{figure}[htbp]
\centering\includegraphics[width=8cm]{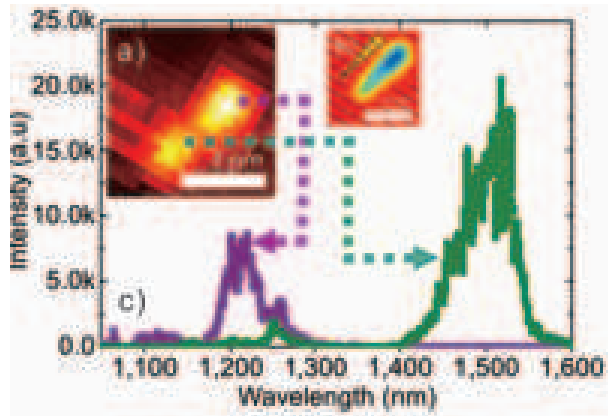}
\caption{(a) PL images of Qdisks in single tapered NW (sample B) on gold substrate. (b) False-color microscope image of the NW with a cartoon for illustrating the NW direction. (c) PL spectra from corresponding PL spots in (a). Measurements were performed using a CW laser at 640 nm wavelength and 4K.}
\label{Fig6}
\end{figure}

Before we performed time-resolved emission from Qdisks, we need to find the origin of the emission peak in the PL spectrum corresponding with the InAs QDisk position. Fig. \ref{Fig3} and Fig. \ref{Fig6} shows the PL scan experiments of a single NW from sample A and B, respectively. Fig. \ref{Fig3}a, b, c, and Fig. \ref{Fig6}a exhibits the PL images of the single NW. Here we put the single NW on gold surface as the PL intensity is much improved by the gold substrate. For this experiment, we set the excitation-beam diameter of $\sim 4\mu$m, shifted the beam position along the NW, and recorded the corresponding PL spectra. Then, the information of the Qdisk $t$ and $d$ can be found by specifying the direction of the NW through the location of the indium particle (Fig. \ref{Fig3}d) or the tapering region (Fig. \ref{Fig6}b) in the scanning electron (SEM) and optical microscope images, respectively. While we put the excitation spot in the NW center of sample A (Fig. \ref{Fig3}b), we observed four emission spots attributed to the Qdisk positions. This is possible because the separation between each Qdisk is about 2 $\mu$m (Fig. \ref{Fig2}a), which is larger than the setup diffraction limit of $\sim\lambda$. On the contrary in sample B, we only observed two emission spot (Fig. \ref{Fig6}a) since the separation between each Qdisk is $\sim$ 500 nm (Fig. \ref{Fig2}b). When we moved the excitation-beam spot to the 2.7-nm and 3.7-nm thick Qdisks in sample A (Fig. \ref{Fig3}a), we observed the strongest peak at the shortest emission wavelength of 1,100-1,200 nm (Fig. \ref{Fig3}e). The emission wavelength shifts towards longer wavelength when we increased $t$ of the Qdisk, {see also Ref. \cite{Zhang2013b} and Appendix A2 for cathodoluminescence and PL-scan measurements, respectively}. In sample B (Fig. \ref{Fig6}a and c), the largest $d$ Qdisk has the shortest emission wavelength. {Here the size of the Qdisk means the diameter of the Qdisk while the thickness of all Qdisks is the same. Due to the large mismatch between InAs and InP, there is large compressive strain in these InAs Qdisks. This compressive strain induces the blueshift of the bandgap. Therefore, the bandgap increases with the diameter of the Qdisk}. Using the information on $d$ and $t$, we can relate $d/\lambda$ with the measured emission rates.

\section{Results}

\begin{figure}[htbp]
\centering\includegraphics[width=7cm]{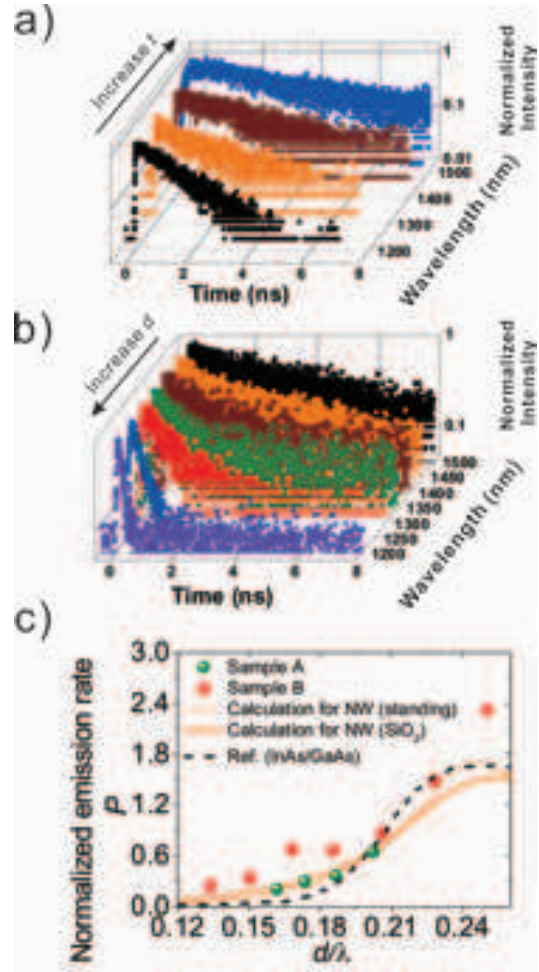}
\caption{(a,b) Time-resolved emission of Qdisks with variation in (a) $t$ (sample A) and (b) $d$ (sample B) inside the single NW on SiO$_{2}$ substrate. (c) Summary of average emission rates. Calculation and reference $P$ from \cite{Bleuse2011} are shown as orange and black lines, respectively. All measurements were performed using a 80-MHz pulsed laser at 800 nm wavelength and 4K.}
\label{Fig5}
\end{figure}

Here we performed time-resolved emission for Qdisks inside single NW on SiO$_{2}$ substrate. Fig. \ref{Fig5}a and b exhibit emission decay curves for Qdisks inside NW as for different $d$ (sample A) and $t$ (sample B), respectively. In sample A, the decay curves were recorded for single NW. For sample B, we collected the decay curves from separated NWs. We fitted all decay curves with single exponential and we found that the emission rates varied from 0.26 to 0.78 ns$^{-1}$ and from 0.30 to 2.78 ns$^{-1}$ for sample A and B, respectively. It is apparent that the emission rate gets smaller towards longer wavelengths in all NWs. Accordingly, the smallest emission rates were recorded for the thickest $t$ and the smallest $d$.

In Fig. \ref{Fig5}c, we summarize the average emission rates as a function of $d/\lambda$. We calculated the average emission rate from five and fourteen NWs for sample A and sample B, respectively. In fact, four data points of sample A were obtained from the sets of emission rates of four Qdisks in five single NWs. The average emission rates in Fig. \ref{Fig5}c were normalized with the emission rate of 1.19 ns$^{-1}$ from the Qdisk in a large NW with $d = 830 \pm 35$ nm and at 1,200 nm emission wavelength, see Appendix {A3}. We expect that the emission rate of 1.19 ns$^{-1}$ is the same as the Qdisk in bulk since the coupling to the guided mode and the screening effect are weak for $d/\lambda > 0.35$ \cite{Bleuse2011}. Additionally, InAs quantum dots in InP bulk wafer shows similar emission rates for the telecommunication emission wavelengths \cite{Birowosuto2012}. {In sample A, the variation of the emission rates is not large as the position of the Qdisk laying on the substrate is always fix and the LDOS influence to the emission rate is averaged over the symmetrical geometry of the Qdisk. For a single quantum dot in the top-down etched NW \cite{Bleuse2011}, the quantum dot position in the NW radial axis is random and thus the fluctuations of the emission rates are large. In case of sample B, the emission rates of the Qdisk fluctuate much larger. We can not explain this trend but the tapering NW has large strain variations, which may change the intrinsic properties from sample to sample.       

Then, we compare our experiments with the calculation of maximum emission rate of InP NW for standing NW and that for NW on SiO$_{2}$ substrate, see orange dashed and orange solid lines in Fig. \ref{Fig5}c, respectively. In general, $d/\lambda$-dependent emission rates from samples A and B can be explained using both calculations for standing NW and NW on SiO$_{2}$ though both samples were laying on SiO$_{2}$. It seems that the SiO$_{2}$ substrate can not alter the inhibition inside the NW. Additionally, in Fig. \ref{Fig1}a, b d, and e, the screening of radiative modes for NW inside air and NW on SiO$_{2}$ substrate are not much different. Thus, there is no influence if the NW is standing or laying on SiO$_{2}$ substrate in regards to the light inhibition or the light confinement inside the NW. This argument confirms that the large diameter bulk NW is still lasing though it is laying on SiO$_{2}$ substrate\cite{Jagadish2013,Mayer2013}. Further investigation on the spontaneous emission rates of the NW will be performed for the NW laying on the high refractive index substrate (i.e. Si).}

\begin{figure}[htbp]
\centering\includegraphics[width=12cm]{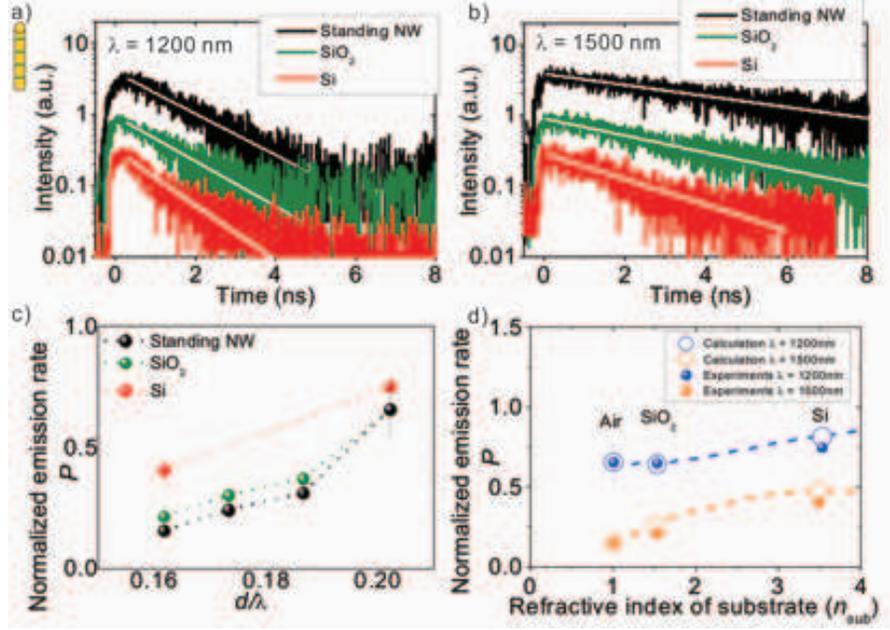}
\caption{(a,b) Time-resolved emission of Qdisks inside the single NW (sample A) for different situations, standing of the grown substrate, laying on SiO$_{2}$, and Si substrates. The emission in (a) and (b) was filtered at 1,200 nm and 1,500 nm, respectively. (c) Collections of the average emission rates from single NW from five NWs in each situation. (d) Normalized decay rates $P$ (obtained from experiments and calculations) as a function of substrate refractive index $n_{sub}$. All measurements were performed using a 80-MHz pulsed laser at 800 nm wavelength and 4K.}
\label{Fig4}
\end{figure}

Fig. \ref{Fig4}a and b shows the time-resolved emission of Qdisk NW (sample A) from standing NW to NW on different substrates at 1,200 and 1,500 nm emission wavelength. Each set of decay curves for standing NW, NW on SiO$_{2}$, and NW on Si was measured from single NW and they were fitted with single exponential. At 1,200 nm emission wavelength (Fig. \ref{Fig4}a), the decay curves are almost the same and the emission rates only vary from 0.74, 0.78, and 0.88 ns$^{-1}$ for standing NW, NW on SiO$_{2}$, and NW on Si, respectively. {Decay curve at 1,500 nm emission wavelength (Fig. \ref{Fig4}b) of NW on Si substrate is completely different than those of NW on SiO$_{2}$ and standing NW. The emission rates are 0.18, 0.26, and 0.42 ns$^{-1}$ for standing NW, NW on SiO$_{2}$, and NW on Si, respectively}. From the strongest inhibition of the emission rates (standing NW), we may estimate that the quantum efficiency of Qdisks is $\gtrsim 76 \%$. Then, we collected the emission rates from five NWs in each situation, i.e. standing and laying on SiO$_{2}$ and Si substrates, and we normalized them with the emission rate of 1.19 ns$^{-1}$ (Fig. \ref{Fig4}c). All three conditions exhibit the same trend as the calculation in Fig. \ref{Fig5}c as the emission rate is strongly suppressed when $d/\lambda$ at 0.16 (1,500 nm). However, at the first sight, the emission inhibition for the NW on Si substrate is less at $d/\lambda$ = 0.16 in comparison with other situations. Thus, to clarify it, we plot the normalized emission rates as a function of substrate refractive indices ($n_{int}$) as shown in Fig. \ref{Fig4}d. Note that $n_{int} = 1$ is the case for standing NW (inside air). Comparing experimental emission rates of standing NW and NW on Si substrate, the inhibition factor {$\zeta$} at $\lambda$ = 1,200 nm decreases from 1.5 to 1.3 while {$\zeta$} at $\lambda$ = 1,500 nm decreases from 6.4 to 2.5. Experimental emission rates well agree with the calculation performed for two $\lambda$. {It means that the decrease of the inhibition for the NW on the Si substrate can be explained by the leaking of the guided and the continuum of radiative modes to the Si substrate as it is shown in the early part of this paper.}

\section{Conclusion}
We have observed the spontaneous emission inhibition of telecom-band Qdisks inside the NW on the arbitrary of the substrate. We controlled the emission rate and the emission wavelength through the different $t$ and $d$ of the multiple Qdisk in the NW. Then, we observed the emission rates as a function $d/\lambda$ and they well agree with calculations considering the emission coupling to the guided and the continuum of the radiative modes. For SiO$_{2}$, the impact of the substrate is relatively small so that the emission rate inhibition is almost the same as that for the standing NW case. For higher-refractive-index substrate, Si, the emission rate inhibition {$\zeta$} at 1,500 nm is about 2.5-fold smaller that for the standing NW. This observation is therefore appealing for designing photonic nanodevices with dielectric platforms. The impact of the substrate to the guided mode and the continuum of radiative modes can be estimated for the efficient devices, such as single photon sources, light emitting devices, and lasers.

\section*{Acknowledgments}
G. Z. would like to acknowledge the support by Scientific Research grant (No: 23310097) from the Japanese Society for the Promotion of Science.

\section*{Appendix}
{
Appendix can be found in the published paper:
Optics Express, Vol. 22, Issue 10, pp. 11713-11726 (2014)
}

\end{document}